\documentstyle[prd,aps,preprint,tighten,floats]{revtex}
\begin{document}
\preprint{CERN-TH/99-149, hep-th/9905155}
\date{May, 1999}
\title{Supergravity Solutions for BI Dyons}
\author{Donam Youm
\thanks{E-mail address: Donam.Youm@cern.ch}}
\address{Theory Division, CERN, CH-1211, Geneva 23, Switzerland}
\maketitle
\begin{abstract}
We construct partially localized supergravity counterpart solutions to 
the 1/2 supersymmetric non-threshold and the 1/4 supersymmetric threshold 
bound state BI dyons in the $D3$-brane Dirac-Born-Infeld theory.  Such 
supergravity solutions have all the parameters of the BI dyons.  By 
applying the IIA/IIB $T$-duality transformations to these supergravity 
solutions, we obtain the supergravity counterpart solutions to  
1/2 and 1/4 supersymmetric BIons carrying electric and magnetic 
charges of the worldvolume $U(1)$ gauge field in the Dirac-Born-Infeld 
theory in other dimensions.  
\end{abstract}
\pacs{11.27.+d, 04.20.Jb, 04.50.+h, 11.25.Sq}

\section{Introduction}

In the very weak string coupling limit ($g_s\to 0$), the worldvolume 
theory of a D-brane is described by the Dirac-Born-Infeld (DBI) 
action \cite{bi,born,clny,dlp,lei} in the flat target spacetime with vanishing 
form and dilaton fields.  Many of solitons in the DBI theory, collectively 
called BIons or BI solitons, have been studied, following the initial works 
in Refs. \cite{gib1,cm,hlw}.  BIons can be regarded as the ``boundary'' 
counterparts to the ``bulk'' supergravity solutions in the bulk/boundary 
holographic duality relations 
\cite{duff1,hoof,town1,suss1,town2,suss2,mald1,mald2,agmoo}.  
More precisely, the intersection or the end (of one brane on the other) 
of an intersecting ``bulk'' supergravity brane solution is interpreted 
in the ``boundary'' worldvolume theory as a soliton in the worldvolume 
theory \cite{pt}.  Such intersection or end is the source for the charge 
carried by the BI soliton \cite{str,towns,town3}.  In fact, it is observed 
\cite{gib1,cm} that in the presence of a BIon, which carries electric 
[magnetic] charge of the worldvolume $U(1)$ gauge field of the $D3$-brane 
DBI action, a flat $D3$-brane develops a transverse spike, which can be 
interpreted as a fundamental string [$D$-string] ending on the $D3$-brane.  
The Coulomb energy density of such configuration is shown \cite{cm} to 
match the energy density of the fundamental string or the $D$-string, 
supporting such idea.  

If the bulk/boundary holographic duality is correct, then all the 
parameters of the worldvolume theory or soliton have to be mapped 
one-to-one to the bulk theory counterparts.  Generally, BIons are 
parameterized by their charges and their locations in the worldvolume 
space.  Therefore, there have to exist the corresponding localized 
intersecting supergravity solutions where at least one brane (interpreted 
as the BI soliton counterpart) is localized on the worldvolume of the 
other (where the BI soliton lives).  

There have been many efforts to construct localized intersecting brane 
solutions 
\cite{tsey1,tsey2,lp,ett,airv,ity,sm,chw,kl,bgkr,ps,yang,youm,fs,gkmt}.  
But it has been turned out \cite{fs,gkmt} that if one works with the 
most general spacetime metric Ansatz for completely localized intersecting 
branes, the resulting equations of motion to be solved are non-linear 
coupled differential equations, which do not have the general solution 
and are almost impossible to solve.  Such equations of motion were solved 
only perturbatively in the case of two $M2$-branes intersecting at a point 
\cite{gkmt}.  If one instead considers the restricted metric Ansatz, which 
is of the same form as that of the delocalized intersecting brane, the 
consistency of the equations of motion requires one of the branes to be 
delocalized on the other \cite{tsey1,tsey2,lp,yang}.  Even with this 
simplified metric Ansatz with one brane delocalized on the other, convenient 
closed form of solutions are generally not known, but only the solutions in 
terms of the infinite series of special functions can be found \cite{lp,sm}.  
But when one goes to the near horizon limit of the delocalized brane, 
solutions can be expressed in closed form in terms of simple 
elementary functions \cite{ity,chw,kl,bgkr,ps,youm}.  Although such 
solutions are not completely localized and in some cases one has to 
delocalize (or compactify) some of the overall transverse directions for 
the purpose of localizing one brane on the other \cite{youm} (when the two 
constituent branes meet at the overall transverse space \cite{mp}), at 
this point this is the only case in which one can have explicit solutions 
in convenient closed forms.  Perhaps, such partially localized solutions might 
be useful for studying the holographic relations in the restricted cases 
where some of directions of the worldvolume brane configurations are 
delocalized.  Furthermore, such partially localized supergravity solutions 
have all the necessary parameters of the corresponding BI solitons, namely 
the locations and the charges of BIons.  

In this paper, we construct the partially localized supergravity solutions 
which are the ``bulk'' counterparts to various BI solitons in the $D3$-brane 
DBI theory.  Many of the partially localized supergravity solutions that 
are the ``bulk'' counterparts to the other solitons in the brane worldvolumes  
are already presented in Ref. \cite{youm}.  For example, first of all for the 
solitons in $M$-brane worldvolume, the self-dual string \cite{hlw} charged 
under the self-dual 3-form field strength in the $M5$-brane worldvolume 
corresponds to the intersecting $M2$- and $M5$-branes where the $M2$-brane 
is localized at the $M5$-brane worldvolume.  The 3-brane soliton \cite{hlw2} 
in the $M5$-brane worldvolume corresponds to the intersecting two $M5$-branes 
where one of them is localized on the other.  The neutral (with respect to 
the self-dual 3-form field strength) string in the $M5$-brane worldvolume 
\cite{glw} corresponds to the $M$-wave localized in the $M5$-brane 
worldvolume.  The zero-brane (charged with respect to the Hodge-dual of a 
transverse scalar) in the $M2$-brane worldvolume corresponds to the 
intersecting two $M2$-branes where one of them is localized on the other.  
Similarly, the partially localized intersecting branes in ten dimensions 
that are presented in Ref. \cite{youm} can also be interpreted as the 
``bulk'' counterparts to the solitons in the ten-dimensional brane 
worldvolumes.  For example, the supergravity solution for $D0$-brane 
localized at the $D4$-brane is the ``bulk'' counterpart to the BI instanton 
in the worldvolume theory of the $D4$-brane \cite{dou,bgt,ggt}, which is 
the double dimensional reduction of the neutral string in the $M5$-brane 
worldvolume.  The 0-brane [2-brane] soliton on the $NS5$-brane worldvolume 
corresponds to the supergravity solution for $D1$-brane [$D3$-brane] 
localized at the $NS5$-brane.  The 3-brane soliton on the 
$NS5$-brane worldvolume corresponds to the intersecting two $NS5$-branes 
where one of them is localized on the other.  The solitons on the worldvolume 
of the Kaluza-Klein (KK) monopole are those on the worldvolume of the 
$NS5$-brane related through the IIA/IIB $T$-duality \cite{pap}, e.g. the 
0-brane [3-brane] soliton on the KK monopole worldvolume corresponds to 
the supergravity solution for the $D2$-brane [$NS5$-brane] localized on 
the KK monopole.  Furthermore, the structure of worldvolume supersymmetry 
algebra dictates the existence of such worldvolume solitons (charged under 
the central charges of the algebra) and the possible intersection rules 
among the spacetime branes \cite{bgt}.

The paper is organized as follows.  In section 2, we survey various 
BI solitons in the $D3$-brane DBI theory.  In section 3, we construct 
the corresponding supergravity solutions and their IIA/IIB $T$-duality 
transformed solutions. 

\section{Solitons in the Dirac-Born-Infeld Theory}

In this section, we summarize the previous works on solitons in 
the Dirac-Born-Infeld theory, which are relevant to the works in 
the following sections, for the purpose of completeness and fixing 
the notation.  

\subsection{Dirac-Born-Infeld Action}

In the very weak string coupling limit ($g_s\to 0$), a $Dp$-brane moves 
in the flat Minkowski spacetime with trivial field configuration, i.e. 
constant dilaton and vanishing form fields.  In this limit, 
a $Dp$-brane with the worldvolume coordinates $\sigma^{\mu}$ 
($\mu=0,1,...,p$) moving in a $(d+1)$-dimensional target space 
with the coordinates $z^M=(x^{\mu},y^m)$ ($M=0,1,...,d$; $\mu=0,1,...,p$, 
$m=p+1,...,d$) is described by the DBI action:
\begin{equation}
S_{DBI}=-\int d^{p+1}\sigma\sqrt{-{\rm det}(\eta_{MN}\partial_{\mu}
z^M\partial_{\nu}z^N+F_{\mu\nu})},
\label{bifact}
\end{equation}
where $\eta_{MN}$ is the metric for the Minkowskian target space and 
$F_{\mu\nu}=\partial_{\mu}A_{\nu}-\partial_{\nu}A_{\mu}$ is the field 
strength of the worldvolume $U(1)$ gauge field $A_{\mu}$.  

In the `static' gauge, in which worldvolume coordinates $\sigma^{\mu}$ 
are identified with target space coordinates as $\sigma^{\mu}=x^{\mu}$, 
the DBI action (\ref{bifact}) takes the following form:
\begin{equation}
S_{DBI}=-\int d^{p+1}\sigma\sqrt{-{\rm det}(\eta_{\mu\nu}+\partial_{\mu}
y^m\partial_{\nu}y^m+F_{\mu\nu})}.
\label{statbifact}
\end{equation}

Since the bulk $D3$-brane configuration is invariant under the 
$SL(2,{\bf R})$ symmetry of the type-IIB theory, the DBI action for 
the $D3$-brane should also have the $SL(2,{\bf R})$ symmetry, as is 
proven in Refs. \cite{gr1,sch,gr2,tse3,gg,town4,ber,kp,ct}.  In fact, the 
manifestly $SL(2,{\bf R})$ covariant form for the $D3$-brane DBI action 
is constructed \cite{cw}, which we describe in the following.  The 
worldvolume $U(1)$ field strength $F$ and its dual field strength $G$, 
defined as $^*G=\pm 2{{\delta S_{DBI}}\over{\delta F}}$, transform as 
a doublet under the $SL(2,{\bf R})$: ${\bf F}\to \Lambda{\bf F}$ 
(${\bf F}=(\matrix{F&G})^T$, $\Lambda\in SL(2,{\bf R})$).  Furthermore, 
the asymptotic values of the dilaton $e^{\phi}$ and the axion $\chi$ 
can be identified with an $SL(2,{\bf R})$ doublet complex constants 
$U^r$ ($r=1,2$), constrained by ${i\over 2}\epsilon_{rs}U^r\bar{U}^s=1$, 
as $U^2/U^1=\chi_{\infty}+ie^{-\phi_{\infty}}$.  Here, $U$ transforms 
under the $SL(2,{\bf R})$ as $U^T\to U^T\Lambda^{-1}$.  Recall that the 
scalars $e^{\phi}$ and $\chi$ of the type-IIB theory form the coset 
$SL(2,{\bf R})/SO(2)$.  This coset is obtained from the complex 
$SL(2,{\bf R})$ doublet ${\cal U}^r$ by gauging the $U(1)\cong SO(2)$ 
which acts as $U^r\to e^{i{\vartheta}}U^r$.  The combination 
${\cal F}=U^T{\bf F}$ is $SL(2,{\bf R})$ invariant.  The $SL(2,{\bf R})$ 
covariant form of the DBI action is expressed in terms of this 
$SL(2,{\bf R})$ invariant ${\cal F}$ and the 4-form field strength $H$ as
\begin{equation}
S_{D3}=\int d^4x\lambda\sqrt{-g}\left[1+{1\over 2}{\cal F}\cdot
\bar{\cal F}-{1\over{16}}({\cal F}\cdot^{\,*}{\cal F})
(\bar{\cal F}\cdot^{\,*}\bar{\cal F})+{1\over 4}H\cdot H\right],
\label{d3bif}
\end{equation}  
where $\lambda$ is a Lagrange multiplier and $g$ is the determinant of 
the worldvolume metric.  This action is supplemented with the 
following self-duality relation
\begin{equation}
{i\over 2}(^*H)^{\,*}{\cal F}={\cal F}-{1\over 4}({\cal F}\cdot 
^{\,*}{\cal F})(\bar{\cal F}\cdot ^{\,*}\bar{\cal F})^{\,*}\bar{\cal F}.
\label{selfdualrel}
\end{equation}
This relation reduces the number of independent charges of $F$ and 
$G$ to two, which can be identified as charges whose sources are 
ends of fundamental string and $D$-string on the $D3$-brane.  

\subsection{BI Solitons}

In this subsection, we survey various solitons in the DBI theory 
whose corresponding and duality related supergravity solutions 
we will construct in the following section.  

By applying the $SL(2,{\bf R})$ transformation, one can bring  
arbitrary constants $U^r$ to $U^1=1$ and $U^2=i$, corresponding to 
the asymptotic values $e^{\phi_{\infty}}=1$ and $\chi_{\infty}=0$ 
for the type-IIB theory scalars.  Note, this choice of the constants 
$U^i$ is preserved by the $U(1)$ subgroup transformation 
discussed in the previous subsection, just like the $SO(2)\subset 
SL(2,{\bf R})$ transformation leaves the scalar asymptotic values 
$e^{\phi_{\infty}}=1$ and $\chi_{\infty}=0$ intact.  With this choice, 
the complex $U(1)$ field strength takes the form ${\cal F}=F+iG$.  

In this case, a general solution, which preserves 1/4 of the worldvolume 
supersymmetry, has the following form 
\cite{gib1,cm,hlw,ggt,blm,gp}:
\begin{eqnarray}
{\bf D}&=&\nabla\sum^{N_q}_{i=1}{{q_i}\over{|{\bf r}-{\bf x}_i|}}, \ \ \ 
X=\sum^{N_q}_{i=1}{{q_i}\over{|{\bf r}-{\bf x}_i|}},
\cr
{\bf B}&=&\nabla\sum^{N_p}_{i=1}{{p_i}\over{|{\bf r}-{\bf y}_i|}}, \ \ \ 
Y=\sum^{N_p}_{i=1}{{p_i}\over{|{\bf r}-{\bf y}_i|}},
\label{genbisol}
\end{eqnarray}
where $X$ and $Y$ are scalars describing target space coordinates 
perpendicular to the worldvolume, and ${\bf r}=(\sigma^1,\sigma^2,
\sigma^3)$ is the spatial worldvolume coordinates.  Here, ${\bf D}$ 
is the canonical momentum conjugate to ${\bf A}=(A_i)$, i.e. the 
electric induction, and ${\bf B}$ is the magnetic induction ($B_i=
{1\over 2}\epsilon_{ijk}F_{jk}$).  The solution describes electric 
and magnetic BIons located at different points in the worldvolume.  
The bulk interpretation of this solution is that fundamental strings 
stretching in the $X$-direction end on the $D3$-brane at ${\bf x}_i$ 
and $D$-strings in the $Y$-direction end at ${\bf y}_i$.  The parameters 
$q_i$ and $p_i$ are respectively related to charges of the fundamental 
strings and $D$-strings.  

A special case of this general solution which is also BPS is the case 
when $N:=N_q=N_p$ and ${\bf x}_i={\bf y}_i$.  Such a solution describes 
$N$ BI dyons with charges $(q_i,p_i)$ located at ${\bf x}_i$.  The bulk 
interpretation is $N$ dyonic strings, each with the NS-NS and the  
R-R two-form potential charges related to $(q_i,p_i)$, ending at the 
points ${\bf r}={\bf x}_i$ in the $D3$-brane worldvolume and stretched 
in the directions $q_i\hat{\bf e}_X+p_i\hat{\bf e}_Y$.  Here, 
$\hat{\bf e}_X$ and $\hat{\bf e}_Y$ are respectively the unit vectors in 
the $X$- and $Y$-directions.  In this case, $1/4$ of spacetime supersymmetry 
is preserved.  The particular case in which both electric and magnetic BIons 
have two centers is interpreted as the string junction \cite{gkmtz}. 
This can be seen by studying the solution in the three asymptotic regions 
near the two centers (${\bf r}\to {\bf x}_1$ and ${\bf r}\to {\bf x}_2$) 
of the BIons and far way from the BIons ($|{\bf r}|\to \infty$).  
The angles between strings in the string junction are determined by the 
charges $q_1$, $q_2$, $p_1$ and $p_2$ of the BIons, and are consistent with 
charge conservation and tension balance, necessary for the existence of the 
static BPS string junction configuration \cite{sch2,dm,sen}.   

The BI soliton that corresponds to the $1/2$ BPS type-IIB dyonic 
string of Schwarz \cite{sch} can be similarly constructed \cite{gp}.  
One starts with the special case of the general solution (\ref{genbisol}) 
in which $p_i=0$ and only one of $q_i$ is non-zero, say $q_1=q
\Delta^{1/2}_{(m,n)}$ and ${\bf x}_1={\bf r}_0$.  Here, we redefined 
the charge with an arbitrary multiplicative constant $\Delta_{(m,n)}$ 
for the purpose of charge quantization after the $SL(2,{\bf R})$ 
transformation.  This corresponds to a fundamental string ending on 
$D3$-brane.  In order to get the BI dyon with electric and magnetic 
charges $(mq,nq)$ and an arbitrary scalar asymptotic value ${\tau}_{\infty}=
\chi_{\infty}+ie^{-\phi_{\infty}}$, one applies the $SL(2,{\bf R})$ 
transformation with the following matrix:
\begin{equation}
\Gamma=\left(\matrix{e^{-{{\phi_{\infty}}\over 2}}&\chi_{\infty}
e^{{\phi_{\infty}}\over 2}\cr 0&e^{{\phi_{\infty}}\over 2}}\right)
\left(\matrix{\cos\alpha&-\sin\alpha\cr \sin\alpha&\cos\alpha}\right),
\label{sl2rbidi}
\end{equation}
with the $SO(2)$ rotation angle $\alpha$ constrained to satisfy:
\begin{equation}
\cos\alpha=e^{{\phi_{\infty}}\over 2}(m-n\chi_{\infty})
\Delta^{-{1\over 2}}_{(m,n)},\ \ \ 
\sin\alpha=e^{-{{\phi_{\infty}}\over 2}}n\Delta^{-{1\over 2}}_{(m,n)}; 
\ \ \ \  m, n \in {\bf Z},
\label{angconst}
\end{equation}
which fix the expression for $\Delta_{(m,n)}$ as:
\begin{equation}
\Delta_{(m,n)}=n^2e^{-\phi_{\infty}}+(n\chi_{\infty}-m)^2
e^{\phi_{\infty}}.
\label{alphaval}
\end{equation}
The resulting dyonic solution has the following form:
\begin{equation}
{\bf D}=\nabla{mq\over{|{\bf r}-{\bf r}_0|}}, \ \ \ \ 
{\bf B}=\nabla{nq\over{|{\bf r}-{\bf r}_0|}}, \ \ \ \ 
X={{\Delta^{1/2}_{(m,n)}q}\over{|{\bf r}-{\bf r}_0|}}.
\label{bidyoinsol}
\end{equation}
This solution is interpreted as the Schwarz dyonic string \cite{sch} 
with the charges related to $(mq,nq)$ and stretching in the $X$-direction 
ending on $D3$-brane.  

\section{Supergravity Solutions}

In this section, we construct the partially localized supergravity 
solution counterparts to the BI solitons in the $D3$-brane worldvolume 
discussed in the previous section. 

\subsection{Type-IIB Supergravity and the $SL(2,{\bf R})$ Symmetry}

In this subsection, we summarize the bosonic effective supergravity 
action for the type-IIB superstring and its $SL(2,{\bf R})$ symmetry 
for the completeness and for the purpose of fixing the notation.

The bosonic part of the low energy effective supergravity theory for 
the type IIB string theory is described by the massless bosonic string 
modes.  These are the graviton $G^{str}_{MN}$, the 2-form potential 
$B^{(1)}_{MN}$ and the dilaton $\phi$ in the NS-NS sector, and the 
axionic scalar $\chi$, the 2-form potential $B^{(2)}_{MN}$ and the 
4-form potential $D_{MNPQ}$ in the R-R sector.  
In the string frame, the bosonic part of the effective supergravity 
action has the following form \cite{bbo}:
\begin{eqnarray}
S^{str}_{IIB}&=&{1\over 2}\int d^{10}x\sqrt{-G^{str}}
\left[e^{-2\phi}\left\{-{\cal R}^{str}+4(\partial\phi)^2
-{3\over 4}(H^{(1)})^2\right\}\right.
\cr
& &\left.-{1\over 2}(\partial\chi)^2-{3\over 4}(H^{(2)}-\chi H^{(1)})^2
-{5\over 6}F^2-{1\over{96\sqrt{-G^{str}}}}\epsilon^{ij}\epsilon
DH^{(i)}H^{(j)}\right],
\label{iibsgact}
\end{eqnarray}
where $H^{(i)}=\partial B^{(i)}$ ($i=1,2$) and $F=\partial D+{3\over 4}
\epsilon^{ij}B^{(i)}B^{(j)}$ are respectively the field strengths of 
the potentials $B^{(i)}$ and $D$.

The $SL(2,{\bf R})$ symmetry \cite{sch3,hw} of the type-IIB theory is manifest 
explicitly in the effective action in the Einstein frame.  By applying the 
Weyl-scaling transformation of spacetime metric, $G_{MN}=e^{-{1\over 2}
\phi}G^{str}_{MN}$, one can bring the action (\ref{iibsgact}) to the 
following Einstein frame form \cite{bbo}:
\begin{eqnarray}
S^{E}_{IIB}&=&{1\over 2}\int d^{10}x\sqrt{-G}\left[-{\cal R}+{1\over 4}
{\rm Tr}(\partial_M{\cal M}\partial^M{\cal M})-{3\over 4}H^{(i)}{\cal M}_{ij}
H^{(j)}\right.
\cr
& &\left.-{5\over 6}F^2-{1\over{96\sqrt{-G}}}\epsilon^{ij}\epsilon 
DH^{(i)}H^{(j)}\right],
\label{iibeinact}
\end{eqnarray}
where an $SL(2,{\bf R})$ invariant matrix ${\cal M}$ formed by the 
scalar $\lambda\equiv \chi+ie^{-\phi}$ has the following form:
\begin{equation}
{\cal M}=e^{\phi}\left(\matrix{|\lambda|^2&\chi\cr \chi& 1}\right).
\label{sl2rscal}
\end{equation}
The Einstein frame action (\ref{iibeinact}) is invariant under the 
following $SL(2,{\bf R})$ transformation:
\begin{equation}
\left(\matrix{B^{(1)}\cr B^{(2)}}\right)\to 
(\Lambda^T)^{-1}\left(\matrix{B^{(1)}\cr B^{(2)}}\right),\ \ \ 
{\cal M}\to \Lambda{\cal M}\Lambda^T;\ \ \ \ 
\Lambda\in SL(2,{\bf R}),
\label{sl2rtr}
\end{equation}
with the metric $G_{MN}$ and the 4-form potential $D$ remaining intact.

\subsection{Supergravity Solution for the Non-Threshold Bound State BI Dyon}

To construct the supergravity counterpart solution to the 1/2 supersymmetric 
BI dyon solution in Eq. (\ref{bidyoinsol}), we start with the following 
partially localized supergravity solution for the fundamental string ending on
\footnote{Strictly speaking, the following solution corresponds to the 
fundamental string ``piercing through'' D3-brane, rather than ending on 
D3-brane.  But at this point, this solution is the closest supergravity 
solution that we have.} 
$D3$-brane \cite{youm}:
\begin{eqnarray}
G^{str}_{MN}dx^Mdx^N&=&-H^{-1}_FH^{-{1\over 2}}_3dt^2
+H^{-{1\over 2}}_3(dx^2_1+dx^2_2+dx^2_3)+H^{-1}_FH^{1\over 2}_3dy^2
\cr
& &+H^{1\over 2}_3(dz^2_1+\cdots+dz^2_5),
\cr
e^{\phi}&=&H^{-{1\over 2}}_F,\ \ \ \ 
B^{(1)}_{ty}=H^{-1}_F,\ \ \  D_{tx_1x_2x_3}=H^{-1}_3,
\label{f3d3sol}
\end{eqnarray}
where the harmonic functions $H_F$ and $H_3$ respectively associated 
with the fundamental string and the $D3$-brane are given by
\footnote{The coupled differential equations that these harmonic functions 
satisfy have the $\delta$-function source terms, which correspond to the 
(microscopic brane) sources to the brane charges.  Namely, strictly speaking 
the coupled differential equations, given in Ref. \cite{youm}, that the 
harmonic functions $H_F$ and $H_3$  satisfy have to be of the following forms:
\begin{equation}
\partial^{2}_{\vec{z}}H_{F}+H_{3}\partial^{2}_{\vec{x}}H_{F}=
q_F\delta({\vec{x}-\vec{x}_0})\delta(\vec{z}-\vec{z}_0), 
\ \ \ 
\partial^2_{\vec{z}}H_3={{q_3}\over{V_y}}\delta({\vec{z}-\vec{z}_0}),
\end{equation}
where $q_3$ and $q_F$ are respectively charges of the fundamental string 
and the D3-brane and $V_y$ is the volume of the delocalized $y$ space.  
Similarly, the differential equations in Eq. (\ref{cnstrdiffeq}) have to be 
modified by the $\delta$-function source terms.  Although the partially 
localized supergravity solutions presented in this paper are valid only 
in the near-horizon region of one of the constituents, one can still define 
charges of the constituent branes through the Gauss law by doing volume 
integrals over the spaces that contain the $\delta$-function singularities 
of the brane charge sources.}
\begin{equation}
H_F=1+{{Q_F}\over{[|\vec{x}-\vec{x}_0|^2+4Q_3|\vec{z}-\vec{z}_0|]^3}}, 
\ \ \ \ 
H_3={{Q_3}\over{|\vec{z}-\vec{z}_0|}}.
\label{f1d3harm}
\end{equation}
Note, in the above two overall transverse directions are delocalized  
(i.e. $\vec{z}$ and $\vec{z}_0$ in the above harmonic functions are 
3-vectors) so that the fundamental string can be localized along the 
worldvolume directions of the $D3$-brane.  

In the Einstein-frame, in which the $SL(2,{\bf R})$ symmetry is 
explicitly manifest in the type-IIB supergravity action, the spacetime 
metric takes the following form:
\begin{eqnarray}
G_{MN}dx^Mdx^N&=&e^{-{1\over 2}\phi}G^{str}_{MN}dx^Mdx^N
\cr
&=&-H^{-{3\over 4}}_FH^{-{1\over 2}}_3dt^2+H^{1\over 4}_FH^{-{1\over 2}}_3
(dx^2_1+dx^2_2+dx^2_3)+H^{-{3\over 4}}_FH^{1\over 2}_3dy^2
\cr
& &+H^{1\over 4}_FH^{1\over 2}_3(dz^2_1+\cdots+dz^2_5).
\label{einf1f3sol}
\end{eqnarray}

In order to construct the supergravity solution corresponding to the 
1/2 BPS BI dyon, we apply the $SL(2,{\bf R})$ $S$-duality transformation 
of the type-IIB supergravity.  We follow the prescription discussed 
in Ref. \cite{sch}.  Before one applies the $SL(2,{\bf R})$ transformation 
(\ref{sl2rtr}) to the supergravity solution (\ref{f3d3sol}) for the 
fundamental string ending on the $D3$-brane, one has to first modify 
the charge $Q_F$ of the fundamental string by multiplying it with an arbitrary 
constant, i.e. $Q_F\to Q_{(q_1,q_2)}\equiv \Delta^{1\over 2}_{(q_1,q_2)}
Q_F$, for the purpose of recovering the charge quantization condition 
\cite{zwa1,zwa2,schw1,schw2,wy1,wy2,wit} after the $SL(2,{\bf R})$ 
transformation.  

We choose the following $SL(2,{\bf R})$ matrix for the transformations:
\begin{eqnarray}
\Lambda&=&\Lambda_1\Lambda_2=\left(\matrix{e^{-\phi_{\infty}/2}&
\chi_{\infty}e^{\phi_{\infty}/2}\cr 0&e^{\phi_{\infty}/2}}\right)
\left(\matrix{\cos\alpha&-\sin\alpha\cr \sin\alpha&\cos\alpha}\right)
\cr
&=&\left(\matrix{e^{-\phi_{\infty}}\cos\alpha+\chi_{\infty}\sin\alpha&
-e^{-\phi_{\infty}}\sin\alpha+\chi_{\infty}\cos\alpha\cr 
\sin\alpha&\cos\alpha}\right)e^{\phi_{\infty}/2}.
\label{sl2rmat}
\end{eqnarray}
The transformation by an $SO(2)$ matrix $\Lambda_2$, which is the most 
general $SL(2,{\bf R})$ matrix that preserves the asymptotic value 
${\cal M}_{\infty}=I$ (or $\lambda_{\infty}=i$) of the scalar matrix 
${\cal M}$, on the solution (\ref{f3d3sol}) results in the supergravity 
solution for a non-threshold bound state of the fundamental string and 
the $D$-string ending on the $D3$-brane with the asymptotic value 
${\cal M}_{\infty}=I$ of the scalar ${\cal M}$.  The subsequent 
transformation on the transformed solution by the matrix $\Lambda_1$ 
leads to the solution with an arbitrary asymptotic value ${\cal M}_{\infty}$ 
(or $\lambda_{\infty}=\chi_{\infty}+ie^{-\phi_{\infty}}$) for the scalar 
fields.  

By demanding that the $SL(2,{\bf R})$ transformed charges $Q^{(1)}$ and 
$Q^{(2)}$ for the fundamental string and the $D$-string to be quantized, i.e. 
\begin{eqnarray}
Q^{(1)}&=&(e^{-{\phi_{\infty}\over 2}}\cos\alpha+\chi_{\infty}e^{\phi_{\infty}
\over 2}\sin\alpha)\Delta^{1/2}_{(q_1,q_2)}Q_F=:q_1Q_F,
\cr  
Q^{(2)}&=&e^{\phi_{\infty}\over 2}\sin\alpha\Delta^{1/2}_{(q_1,q_2)}Q_F
=:q_2Q_F, 
\label{sl2elecchs}
\end{eqnarray}
for any integers $q_1$ and $q_2$, one fixes the $SO(2)$ rotation angle 
$\alpha$ to be
\begin{equation}
\left.\matrix{\cos\alpha=e^{\phi_{\infty}\over 2}(q_1-q_2\chi_{\infty})
\Delta^{-{1\over 2}}_{(q_1,q_2)}\cr
\sin\alpha=e^{-{\phi_{\infty}\over 2}}q_2
\Delta^{-{1\over 2}}_{(q_1,q_2)}}\right\}\ \ \ 
\Leftrightarrow \ \ \ 
e^{i\alpha}=e^{\phi_{\infty}\over 2}(q_1-q_2\bar{\lambda}_{\infty})
\Delta^{-{1\over 2}}_{(q_1,q_2)};\ \ \ \ 
q_1, q_2\in {\bf Z}.
\label{angconstr}
\end{equation}
From this, one determines the expression for the arbitrary constant 
$\Delta_{(q_1,q_2)}$ to be
\begin{equation}
\Delta_{(q_1,q_2)}\equiv e^{\phi_{\infty}}(q_1-q_2\chi_{\infty})^2+
e^{-\phi_{\infty}}q^2_2=\left(\matrix{q_1& q_2}\right){\cal M}^{-1}_{\infty}
\left(\matrix{q_1\cr q_2}\right).
\label{deltaform}
\end{equation}
The $SL(2,{\bf R})$ transformed charges $(Q^{(1)},Q^{(2)})=(q_1Q_F,q_2Q_F)$ 
of the fundamental string and the $D$-string satisfy the charge quantization 
condition.  

The final form of the $SL(2,{\bf R})$ transformed solution in the Einstein 
frame describing the non-threshold bound state of the fundamental 
string and the $D$-string ending on the $D3$-brane with arbitrary 
asymptotic values for the scalars is as follows:
\begin{eqnarray}
G_{MN}dx^Mdx^N&=&-H^{-{3\over 4}}_{(q_1,q_2)}H^{-{1\over 2}}_3dt^2
+H^{1\over 4}_{(q_1,q_2)}H^{-{1\over 2}}_3(dx^2_1+dx^2_2+dx^2_3)
+H^{-{3\over 4}}_{(q_1,q_2)}H^{1\over 2}_3dy^2
\cr
& &+H^{1\over 4}_{(q_1,q_2)}H^{1\over 2}_3(dz^2_1+\cdots+dz^2_5),
\cr
\lambda&=&{{q_1\chi_{\infty}-q_2|\lambda_{\infty}|^2+iq_1e^{-\phi_{\infty}}
H^{1\over 2}_{(q_1,q_2)}}\over
{q_1-q_2\chi_{\infty}+iq_2e^{-\phi_{\infty}}H^{1\over 2}_{(q_1,q_2)}}},
\cr
\left(\matrix{B^{(1)}_{ty}\cr B^{(2)}_{ty}}\right)&=&
{\cal M}^{-1}_{\infty}\left(\matrix{q_1\cr q_2}\right)
\Delta^{-1/2}_{(q_1,q_2)}H^{-1}_{(q_1,q_2)},\ \ \ 
D_{tx_1x_2x_3}=H^{-1}_3.
\label{einf1d1d3sol}
\end{eqnarray}
From the above expression for $\lambda$, one has the following 
forms for the axion $\chi$ and the dilaton $e^{\phi}$:
\begin{eqnarray}
\chi&=&{{e^{\phi_{\infty}}\chi_{\infty}\Delta_{(q_1,q_2)}
+q_1q_2(H_{(q_1,q_2)}-1)}\over{q^2_2H_{(q_1,q_2)}+e^{2\phi_{\infty}}
(q_1-q_2\chi_{\infty})^2}},
\cr
e^{-\phi}&=&{{\Delta_{(q_1,q_2)}H^{1/2}_{(q_1,q_2)}}\over
{q^2_2H_{(q_1,q_2)}+e^{2\phi_{\infty}}(q_1-q_2\chi_{\infty})^2}}
=:e^{-\phi_{\infty}}\bar{H}^{-1}_{(q_1,q_2)}H^{1/2}_{(q_1,q_2)},
\label{axidil}
\end{eqnarray}
where
\begin{equation}
\bar{H}_{(q_1,q_2)}={{e^{-\phi_{\infty}}q^2_2H_{(q_1,q_2)}
+e^{\phi_{\infty}}(q_1-q_2\chi_{\infty})^2}\over{\Delta_{(q_1,q_2)}}}.
\label{defofh}
\end{equation}
Here, the harmonic functions, which describe the non-threshold bound state 
of the fundamental string and the $D$-string localized on the worldvolume 
of the $D3$-brane, are given by 
\begin{eqnarray}
H_{(q_1,q_2)}&=&1+{{Q_{(q_1,q_2)}}\over{[|\vec{x}-\vec{x}_0|^2+4Q_3
|\vec{z}-\vec{z}_0|]^3}}, 
\cr 
\bar{H}_{(q_1,q_2)}&=&1+{{e^{-\phi_{\infty}}q^2_2\Delta^{-1}_{(q_1,q_2)}
Q_{(q_1,q_2)}}\over{[|\vec{x}-\vec{x}_0|^2+4Q_3
|\vec{z}-\vec{z}_0|]^3}},
\cr
H_3&=&{{Q_3}\over{|\vec{z}-\vec{z}_0|}},
\label{sl2harmlocal}
\end{eqnarray}
where $Q_{(q_1,q_2)}=\Delta^{1/2}_{(q_1,q_2)}Q_F$.

Note, this solution has all the parameters of the 1/2 BPS BI dyon, 
namely $\vec{x}_0$ that is related to the location of the BI dyon in the 
$D3$-brane worldvolume and $(q_1,q_2)$ that are related to the electric 
and the magnetic charges of the BI dyon.  Note also that the parameter 
$q\Delta_{(m,n)}$ in the expression for $X$ in Eq. (\ref{bidyoinsol}) is 
related to $Q_{(q_1,q_2)}$ in the above harmonic function $H_{(q_1,q_2)}$ 
for the $(q_1Q_F,q_2Q_F)$-string.  This is related to the fact that the 
magnitude of the scalar charge (of the scalar $X$ in Eq. (\ref{bidyoinsol})) 
can be identified with the tension of the attached string, i.e. the 
above $(q_1Q_F,q_2Q_F)$-string.  The direction along which the 
$(q_1Q_F,q_2Q_F)$-string stretches is the direction associated with the 
worldvolume scalar $X$.   

By applying the IIA/IIB $T$-duality transformations on the solution 
(\ref{einf1d1d3sol}) along the overall transverse directions $z_i$, one 
obtains partially localized supergravity solutions that correspond to 
new class of non-threshold bound state BI solitons.  Before one applies 
the $T$-duality transformations, one has to first delocalize the solution
along those directions, i.e. one has to compactify the $T$-dualized directions 
on circles.  Note, however that two of the overall transverse directions 
of the solution (\ref{einf1d1d3sol}) are already delocalized for the 
purpose of localizing the string on the worldvolume of the $D3$-brane.  
The resulting supergravity solution has the following form:
\begin{eqnarray}
G_{MN}dx^Mdx^N&=&e^{{{n+2}\over 8}\phi_{\infty}}\bar{H}^{{n+2}\over 8}_{1+n}
H^{1\over 4}_1H^{{n-4}\over 8}_{3+n}\left[-H^{-1}_1dt^2+dx^2_1+dx^2_2+dx^2_3
\right.
\cr
& &+e^{-\phi_{\infty}}\bar{H}^{-1}_{1+n}(du^2_1+\cdots+du^2_n)
+H^{-1}_1H_{3+n}dy^2
\cr
& &\left.+H_{3+n}(dz^2_1+\cdots+dz^2_{4-n})\right],
\cr
e^{\phi}&=&e^{{{2-n}\over 4}\phi_{\infty}}\bar{H}^{{2-n}\over 4}_{1+n}
H^{-{1\over 2}}_1H^{-{n\over 4}}_{3+n},
\label{nonthrestdual}
\end{eqnarray}
where in the case of $n=1,2$ the harmonic functions are given by
\begin{eqnarray}
H_1&=&1+{{Q_{(q_1,q_2)}}\over{[|\vec{x}-\vec{x}_0|^2+4Q_3
|\vec{z}-\vec{z}_0|]^3}},
\cr
\bar{H}_{1+n}&=&1+{{e^{-\phi_{\infty}}q^2_2\Delta^{-1}_{(q_1,q_2)}
Q_{(q_1,q_2)}}\over{[|\vec{x}-\vec{x}_0|^2+4Q_3
|\vec{z}-\vec{z}_0|]^3}},
\cr
H_{3+n}&=&{{Q_3}\over{|\vec{z}-\vec{z}_0|}}.
\label{nonthtdualharm}
\end{eqnarray}

The solution describes the non-threshold bound state of the fundamental 
string (with the longitudinal coordinate $y$ and the associated harmonic 
function $H_1$) and the $D(1+n)$-brane (with the longitudinal coordinates 
$(y,u_1,...,u_n)$ and the associated harmonic function $\bar{H}_{1+n}$) 
ending on the $D(3+n)$-brane (with the longitudinal coordinates 
$(x_1,x_2,x_3,u_1,...,u_n)$ and the associated harmonic function $H_{3+n}$).  

From this solution, one can see the existence of the 1/2 BPS non-threshold 
bound state of an electric BIon and a magnetic BIon under the worldvolume 
$U(1)$ gauge field of the $(3+n)$-dimensional DBI theory.  Such magnetic 
BIon is related to the self-dual string \cite{hlw} in the worldvolume 
of the $M5$-brane through the dimensional reduction (along the 
worldvolume direction transverse to the self-dual string) and the IIA/IIB 
$T$-dualities (in the transverse directions).  Also, the electric BIon has 
the same origin in the $M5$-brane worldvolume: for this case, the longitudinal 
direction  of the self-dual string is compactified.   One can think of such 
non-threshold bound states of the electric and magnetic BIons in the 
$D(3+n)$-brane worldvolume as being originated from the self-dual 
string soliton in the $M5$-brane worldvolume wrapped around a ``tilted'' 
circle (at angle with respect to the longitudinal and a transverse 
direction) \cite{cp,hlw}, followed by the IIA/IIB $T$-duality transformations. 

In the string frame, the above solution (\ref{nonthrestdual}) takes more 
recognizable form.  After redefining the scalar asymptotic values such 
that the resulting spacetime metric in the string frame is asymptotically 
flat and then applying the Weyl-rescale transformation of the metric 
$G^{str}_{MN}=e^{\phi/2}G_{MN}$, one obtains the following string frame 
form of the spacetime metric:
\begin{eqnarray}
G^{str}_{MN}dx^Mdx^N&=&-H^{-1}_1\bar{H}^{1\over 2}_{1+n}
H^{-{1\over 2}}_{3+n}dt^2+\bar{H}^{1\over 2}_{1+n}H^{-{1\over 2}}_{3+n}
(dx^2_1+dx^2_2+dx^2_3)
\cr
& &+\bar{H}^{-{1\over 2}}_{1+n}H^{-{1\over 2}}_{3+n}
(du^2_1+\cdots+du^2_n)+H^{-1}_1\bar{H}^{1\over 2}_{1+n}
H^{1\over 2}_{3+n}dy^2
\cr
& &+\bar{H}^{1\over 2}_{1+n}H^{1\over 2}_{3+n}
(dz^2_1+\cdots+dz^2_{4-n}).
\label{strtdualnonth}
\end{eqnarray}

\subsection{Supergravity Solution for the Threshold Bound State BI Dyon}

The ``bulk'' spacetime configuration counterpart to the threshold 
bound state of electric BIons and magnetic BIons is the orthogonal 
fundamental and $D$ strings ending on $D3$-brane with the following 
configuration:
\begin{equation}
\matrix{D3:&1&2&3&-&-&-&-&-&-\cr D1:&-&-&-&4&-&-&-&-&-\cr 
F1:&-&-&-&-&5&-&-&-&-}
\label{f1d1d3config}
\end{equation}
In the string frame, such configuration  is described by the following 
supergravity solution, which is the ``bulk'' counterpart to the general 
BI dyon in Eq. (\ref{genbisol}):
\begin{eqnarray}
G^{str}_{MN}dx^Mdx^N&=&-H^{-1}_FH^{-{1\over 2}}_1H^{-{1\over 2}}_3dt^2+
H^{1\over 2}_1H^{-{1\over 2}}_3(dx^2_1+dx^2_2+dx^2_3)+
H^{-{1\over 2}}_1H^{1\over 2}_3du^2
\cr
& &+H^{-1}_FH^{1\over 2}_1H^{1\over 2}_3dy^2+H^{1\over 2}_1H^{1\over 2}_3
(dz^2_1+\cdots+dz^2_4),
\cr
e^{\phi}&=&H^{-{1\over 2}}_FH^{1\over 2}_1,\ \ \ 
B^{(1)}_{ty}=H^{-1}_F,\ \ \ 
B^{(2)}_{tu}=H^{-1}_1,\ \ \ 
D_{tx_1x_2x_3}=H^{-1}_3,
\label{f1d1d3thres}
\end{eqnarray}
where the harmonic functions satisfy the following coupled differential 
equations:
\begin{eqnarray}
\partial^2_{\vec{z}}H_F+H_3\partial^2_{\vec{x}}H_F+H_1\partial^2_uH_F&=&0,
\cr
\partial^2_{\vec{z}}H_1+H_3\partial^2_{\vec{x}}H_1+H_F\partial^2_yH_3&=&0,
\cr
\partial^2_{\vec{z}}H_3+H_1\partial^2_uH_3+H_F\partial^2_yH_3&=&0,
\label{f1d1d3diffeq}
\end{eqnarray}
along with the constraints
\begin{equation}
\partial_uH_F\partial_yH_p=0,\ \ \ 
\partial_yH_3\partial_{\vec{x}}H_F=0,\ \ \ 
\partial_uH_3\partial_{\vec{x}}H_1=0.
\label{f1d1d3cnstrnt}
\end{equation}

The constraints (\ref{f1d1d3cnstrnt}) can be satisfied when the solution 
(\ref{f1d1d3thres}) is delocalized along $y$ and $u$ directions.  
With this choice, the fundamental strings and the $D$-strings become 
completely localized at the $D3$-brane.  In this case, the coupled 
differential equations (\ref{f1d1d3diffeq}) satisfied by the harmonic 
functions reduce to
\begin{equation}
\partial^2_{\vec{z}}H_F+H_3\partial^2_{\vec{x}}H_F=0,\ \ \ 
\partial^2_{\vec{z}}H_1+H_3\partial^2_{\vec{x}}H_1=0,\ \ \ 
\partial^2_{\vec{z}}H_3=0.
\label{cnstrdiffeq}
\end{equation}
One can solve these coupled differential equations by following the procedure 
discussed in Ref. \cite{youm}.  Since the dimensionality of the overall 
transverse space is 4, one has to delocalize one of the overall transverse 
directions in order for the fundamental string and the $D$-string to be 
localized along the worldvolume directions of the $D3$-brane.  The 
expressions for the harmonic functions have the 
following forms:
\begin{eqnarray}
H_F&=&1+\sum^{N_F}_{i=1}{{Q_{F\,i}}\over{[|\vec{x}-\vec{x}_{F1\,i}|^2
+4Q_3|\vec{z}-\vec{z}_0|]^{3\over 2}}},
\cr
H_1&=&1+\sum^{N_D}_{i=1}{{Q_{D\,i}}\over{[|\vec{x}-\vec{x}_{D1\,i}|^2
+4Q_3|\vec{z}-\vec{z}_0|]^{3\over 2}}},
\cr
H_3&=&{{Q_3}\over{|\vec{z}-\vec{z}_0|}}.
\label{dfd1d3localharm}
\end{eqnarray}
Note, these ``localized'' harmonic functions contain all the parameters 
of the electric BIons and the magnetic BIons in the 
worldvolume of the $D3$-brane.  Namely, the electric and magnetic charges 
$q_i$ and $p_i$ of the BIons (\ref{genbisol}) are related to the 
charges $Q_{F\,i}$ and $Q_{D\,i}$ of the fundamental strings and the $D$ 
strings, and the locations ${\bf x}_i$ and ${\bf y}_i$ of the electric 
BIons and the magnetic BIons are related to the locations $\vec{x}_{F1\,i}$ 
and $\vec{x}_{D1\,i}$ of the fundamental strings and the $D$ strings in 
the $D3$-brane worldvolume directions. 

As in the case of the threshold bound state BI dyon solution 
(\ref{genbisol}), when $N_F=N_D$ and $\vec{x}_{F1\,i}=\vec{x}_{D1\,i}$ 
the solution (\ref{f1d1d3thres}) describes the threshold bound state 
of dyonic strings with the NS-NS and R-R two-form charges 
$(Q_{F\,i},Q_{D\,i})$ stretched in the $Q_{F\,i}\hat{\bf e}_y+Q_{D\,i}
\hat{\bf e}_u$ directions.  Here, $\hat{\bf e}_y$ and $\hat{\bf e}_u$ are 
respectively the unit vectors in the $y$ and $u$ directions.  Since these 
dyons are at angle with respect to the $y$ and $u$ coordinates, there has 
to be non-zero off-diagonal term $G^{str}_{uy}$ indicating rotation of the 
dyonic strings in the $uy$-plane.  But the above supergravity solution 
lacks such term, because the solution is delocalized along those directions. 
Such off-diagonal term is expected to appear in the fully localized 
solution with more general metric Ansatz.  
The particular case in which both of the harmonic functions $H_F$ and $H_1$ 
have two centers corresponds to the ``bulk'' spacetime counterpart 
configuration to the string junction discussed in the previous section.  
As a string approaches the $D3$-brane, it splits into two strings.  

By applying the IIA/IIB $T$-duality transformations along the overall 
transverse directions of the solution (\ref{f1d1d3thres}), one obtains 
the following supergravity solution for the fundamental string and the 
$Dp$-brane ending on the $D(p+2)$-brane:
\begin{eqnarray}
G^{str}_{MN}dx^Mdx^N&=&-H^{-1}_FH^{-{1\over 2}}_pH^{-{1\over 2}}_{p+2}dt^2 
+H^{-{1\over 2}}_pH^{-{1\over 2}}_{p+2}(dw^2_1+\cdots+dw^2_{p-1})
\cr
& &+H^{1\over 2}_pH^{-{1\over 2}}_{p+2}(dx^2_1+dx^2_2+dx^2_3)
+H^{-{1\over 2}}_pH^{1\over 2}_{p+2}du^2
\cr
& &+H^{-1}_FH^{1\over 2}_pH^{1\over 2}_{p+2}dy^2
+H^{1\over 2}_pH^{1\over 2}_{p+2}(dz^2_1+\cdots+dz^2_{5-p}),
\cr
e^{\phi}&=&H^{-{1\over 2}}_FH^{{3-p}\over 4}_pH^{{1-p}\over 4}_{p+2}, 
\cr
B^{(1)}_{ty}&=&H^{-1}_F,\ \ \  A_{tw_1\cdots w_{p-1}u}=H^{-1}_p,\ \ \ 
A_{tw_1\cdots w_{p-1}x_1x_2x_3}=H^{-1}_{p+2}.
\label{tdualf1d1d3thres}
\end{eqnarray}

In the case of $p=2$ (the fundamental string and the $D2$-brane ending on 
the $D4$-brane), the harmonic functions are given by
\begin{eqnarray}
H_F&=&1+\sum_i{{Q_{F\,i}}\over{[|\vec{x}-\vec{x}_{F1\,i}|^2
+4Q_4|\vec{z}-\vec{z}_0|]^{3\over 2}}},
\cr
H_2&=&1+\sum_i{{Q_{2\,i}}\over{[|\vec{x}-\vec{x}_{D2\,i}|^2
+4Q_4|\vec{z}-\vec{z}_0|]^{3\over 2}}},
\cr
H_{4}&=&{{Q_{4}}\over{|\vec{z}-\vec{z}_0|}}.
\label{f1d2d4localharm}
\end{eqnarray}
From this supergravity solution, one infers that there should exist 
the 1/4 BPS threshold bound state of an electric BIon and a magnetic 
BIon under the worldvolume $U(1)$ gauge field of the 4-dimensional 
DBI theory.  This solution is originated from the two intersecting 
self-dual strings on the $M5$-brane worldvolume \cite{glw} through 
the dimensional reduction along the longitudinal direction of one 
of the self-dual strings.  The corresponding ``bulk'' spacetime 
counterpart configuration is
\begin{equation}
\matrix{M5:&1&2&3&4&5&-&-&-&-&-\cr M2:&-&-&-&4&-&6&-&-&-&-\cr 
M2:&-&-&-&-&5&-&7&-&-&-}
\label{twoselfdualstr}
\end{equation}
The two worldvolume self-dual strings are respectively stretched along 
the 4 and 5 directions of the $M5$-brane worldvolume.  
After compactifying either 4 or 5 on a circle, one obtains the 
above ten-dimensional configuration describing the fundamental 
string and the $D2$-brane ending on the $D4$-brane.  The corresponding 
worldvolume configuration is the threshold bound state of the (electric) 
0-brane and the (magnetic) 1-brane in the $D4$-brane DBI theory.  
The supergravity solution for the fundamental string and the $Dp$-brane 
ending on the $D(p+2)$-brane with $p>2$, which is related to the 
configuration (\ref{twoselfdualstr}) through the dimensional reduction 
and the IIA/IIB $T$-dualities, also implies existence of the 1/4 BPS threshold 
bound state of electric BIon (0-brane) and a magnetic BIon ($(p-1)$-brane) 
under the worldvolume $U(1)$ gauge field of the $(p+3)$-dimensional DBI 
theory of the $D(p+2)$-brane.

\end{document}